\documentclass[
    11pt,
    letterpaper,
    reprint,
    nofootinbib,
    notitlepage,
    superscriptaddress,
    aps,
    physrev,
]{revtex4-2}

\usepackage{amsmath,amssymb,amsfonts, amsthm} 
\usepackage{physics}

\usepackage{stackengine,graphicx}
\usepackage{etaremune}
\usepackage{braket}
\usepackage{euscript}
\usepackage{placeins}
\usepackage{mathtools}
\usepackage{mathrsfs}
\usepackage{appendix}
\usepackage{soul} 

\usepackage{xy}
\xyoption{matrix}
\xyoption{frame}
\xyoption{arrow}
\xyoption{arc}

\usepackage{ifpdf}
\ifpdf
\else
\PackageWarningNoLine{Qcircuit}{Qcircuit is loading in Postscript mode.  The Xy-pic options ps and dvips will be loaded.  If you wish to use other Postscript drivers for Xy-pic, you must modify the code in Qcircuit.tex}
\xyoption{ps}
\xyoption{dvips}
\fi

\entrymodifiers={!C\entrybox}

\usepackage{algorithm}
\usepackage{algpseudocode}
\usepackage[svgnames,dvipsnames]{xcolor}

\definecolor{NewBlue}{rgb}{0.1, 0.1, 0.7}
\definecolor{NewRed}{rgb}{0.7, 0.1, 0.1}

\usepackage[colorlinks,
    linkcolor=Maroon,
    citecolor=NewBlue,
    urlcolor=NewRed]{hyperref}
    
\usepackage{cleveref}

\begin{document}

\title{Fault Tolerant Non-Clifford State Preparation for Arbitrary Rotations}
\author{Hyeongrak Choi}
\email[]{choihr@mit.edu}
\affiliation{Research Laboratory of Electronics, Massachusetts Institute of Technology, Cambridge, Massachusetts 02139, USA}
\author{Frederic T. Chong}
\email[]{chong@cs.uchicago.edu}
\affiliation{Computer Science, University of Chicago, Chicago, Illinois 60615, USA}
\author{Dirk Englund}
\email[]{englund@mit.edu}
\affiliation{Research Laboratory of Electronics, Massachusetts Institute of Technology, Cambridge, Massachusetts 02139, USA}
\author{Yongshan Ding}
\email[]{yongshan.ding@yale.edu}
\affiliation{Computer Science, Yale University, New Haven, Connecticut 06511, USA}
\affiliation{Yale Quantum Institute, Yale University, New Haven, Connecticut 06520, USA}

\date{\today}

\begin{abstract}
Quantum error correction is an essential component for practical quantum computing on noisy quantum hardware. However, logical operations on error-corrected qubits require a significant resource overhead, especially for high-precision and high-fidelity non-Clifford rotation gates. To address this issue, we propose a postselection-based algorithm to efficiently prepare resource states for gate teleportation. Our algorithm achieves fault tolerance, demonstrating the exponential suppression of logical errors with code distance, and it applies to any stabilizer codes. We provide analytical derivations and numerical simulations of the fidelity and success probability of the algorithm. We benchmark the method on surface code and show a factor of $10^2$ to $10^4$ reduction in space-time overhead compared to existing methods. Overall, our approach presents a promising path to reducing the resource requirement for quantum algorithms on error-corrected and noisy intermediate-scale quantum computers.
\end{abstract}

\maketitle

\section{Introduction}

Practical quantum computers should reliably store and process quantum information in the presence of noise and error. Quantum error correction (QEC) encodes information into a set of physical qubits to protect them from environmental noises. Moreover, to overcome erroneous physical operations, one needs to apply only fault-tolerant (FT) gates to encoded logical qubits. Transversal gates achieve this by prohibiting errors from propagating within a code block. Unfortunately, it has been proven that there is no code with universal transversal gate set~\cite{gottesman1998heisenberg,eastin2009restrictions}.

An alternative route to FT gates is preparing ancillae in a state not accessible from transversal gates and gate-teleport them~\cite{bravyi2005universal}. Logical states of a code are often constructed with state injection~\cite{li2015magic}. The fidelity of this logical state is limited by the fidelity of the ancilla state regardless of the code distance. Moreover, even for a perfect ancilla state, the single-qubit and two-qubit gate fidelities limit the fidelity of the logical state. This is because the information to be injected is in a physical qubit and needs to pass through the gates for encoding.

For fault-tolerant computing, a conventional approach to prepare an ancilla is magic state distillation~\cite{bravyi2005universal}. In a magic state distillation, error detection is performed on faulty logical qubits with the inner code embedding transversal non-Clifford gate~\cite{bravyi2012magic}. Alternatively, a code with transversal Clifford gates can be used for transversal Clifford measurement~\cite{meier2012magic}. 15-to-1 Reed-Muller code has been widely studied~\cite{knill1996threshold}, but other codes~\cite{bravyi2012magic, meier2012magic} and multi-level distillation~\cite{jones2013multilevel} have been also proposed. 

The overhead of distillation is often significantly higher than that of a Clifford operation. For example, a recent study \cite{litinski2019magic} showed that the surface code implementation of the distillation circuit only requires $\sim 4.71d^3$ $(d=13)$ space-time cost (i.e., qubits multiplied by cycles) for $p_L\sim 10^{-9}$ logical error rate provided physical error rate of $p = 10^{-4}$, which is only $\sim 5$ times more expensive than the Clifford gate taking $d^3$ qubit cycles. However, achieving $p_L < 10^{-10}$ with realistic $p = 10^{-3}$ requires $>30d^3$ qubit-cycles. Moreover, the relative cost is much higher for codes with a higher encoding rate or a lower threshold. Alternatively, transversal $T$-measurement on 2D color code augmented with flag qubits and redundant ancilla has shown a low overhead distillation~\cite{chamberland2020very}. Although it is 2D compatible and can be code-switched to surface code using lattice surgery~\cite{nautrup2017fault}, the low threshold of $0.2 \%$~\cite{chamberland2020triangular} limits its usage to a system with a very low physical error rate of $p\approx 10^{-4}$. 

There have been a few ways to bypass the distillation. 1) Code switching with 3D Gauge color code implements transversal $T$-gate with gauge fixing~\cite{bombin2015gauge, paetznick2013universal}. However, the threshold is much lower ($\sim3\times10^{-3}$ for single-qubit Pauli-$X$ errors) ~\cite{brown2016fault}, it strictly requires a 3D physical connectivity, and the resource overhead is still higher than the distillation \cite{beverland2021cost}. 2) The code deformations of the 2D surface code effectively perform the 3D surface code and enable the transversal CCZ gate~\cite{brown2019fault}. The problem is that this requires local three-dimensional operations for a transversal CCZ of three copies of the surface code patch. Essentially, by the Bravyi-Koenig bound, one needs at least a three-dimensional code for transversal non-Clifford gates \cite{bravyi2013classification}.

In addition, a fault-tolerant implementation of arbitrary-angle rotations requires multiple $T$-gates~\cite{bocharov2015efficient, ross2014optimal, paetznick2013repeat}. Such implementation is practically important because, for example, quantum simulation of 100 orbitals requires $10^{12}$ single-qubit rotations with angle $\theta\approx 10^{-7}$~\cite{poulin2014trotter, hastings2014improving, wecker2014gate}, resulting in a formidable resource overhead. Recently, $T$ count was greatly reduced by parity-checking protocol \cite{duclos2015reducing,campbell2016efficient} and improved by the so-called ``synthilation'' \cite{campbell2018magic}. In synthilation, gate synthesis is optimized in conjunction with distillation in a single step. However, even with quadratic suppression of the $T$-error rate, performing multiple stages of synthilation for arbitrary-angle rotations remains challenging. Moreover, a large number of logical qubits with enough distances unavoidably takes many physical qubits.

The oxymoron is that small-angle rotations, almost the same as Clifford gates, require \emph{more} resources than a $T$ gate, which is farthest from the Clifford octahedron on the Bloch sphere equator. Our work aims to fill this gap using a novel scheme for preparing arbitrary-angle exotic magic states in a resource-efficient way.

Here, we present a method for fault-tolerantly preparing a broad family of non-Clifford states. Our method achieves fault tolerance by taking advantage of post-selection through a quantum \emph{error detecting code}. We demonstrate its fault-tolerance property by showing that the logical error rate decays exponentially with the code distance. We also analyze the commonly ignored coherent errors and show that they can be statistically suppressed. Furthermore, we introduce a technique called ``rotation scaffolding'' to boost the success rate of a logical rotation gate. As a result, the protocol reduces the space-time overhead of preparing non-Clifford states by \emph{several orders of magnitude}. Compared to existing methods, our method uses a factor of $10^2$ to $10^4$ lower space-time cost to achieve a logical error rate of $10^{-8}$ to $10^{-4}$. This is especially useful for bootstrapping NISQ devices into near-term FT systems. The generated states can be fed into other existing distillation protocols for even lower error rates. 

\begin{figure*}
    \centering
    \includegraphics[width=\textwidth]{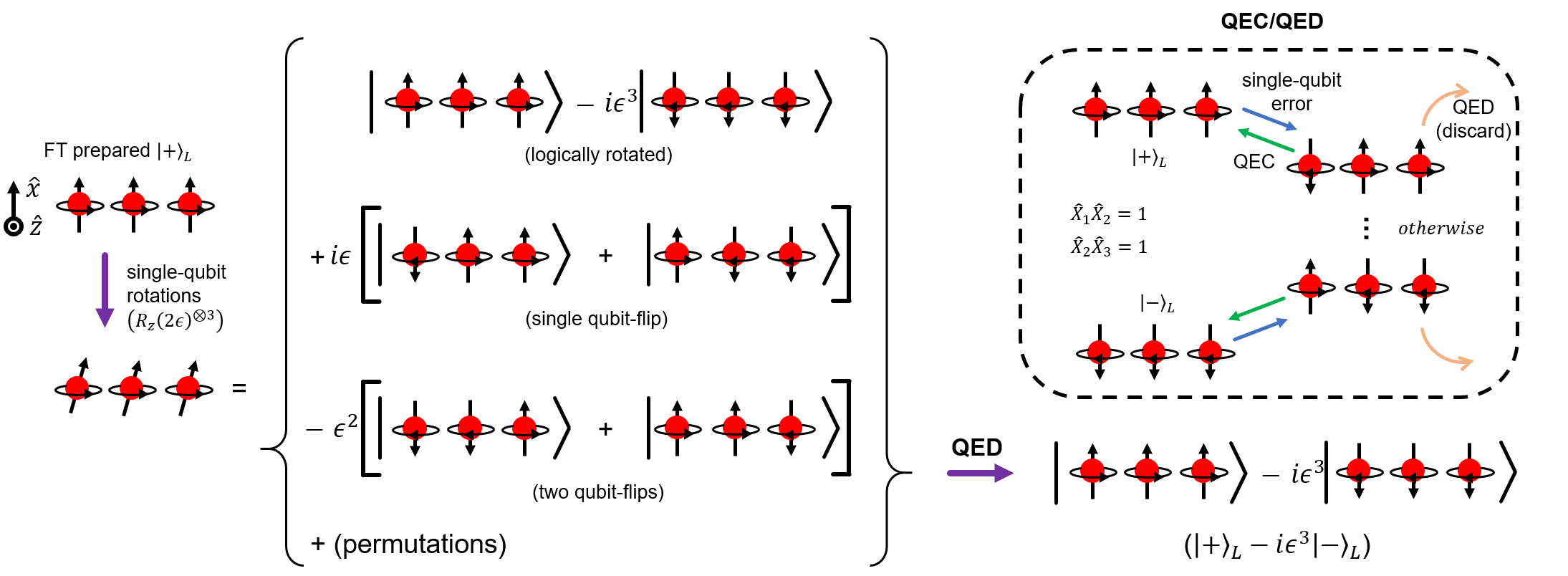}
    \caption{\textbf{Fault-tolerant non-Clifford state preparation}. Description of our method for the three-qubit phase-flip code. First, the logical qubit is prepared in the Clifford state, $\ket{+_L} = \ket{+++}$. Note that this is different from the conventional definition, $\ket{0_L} = \ket{+++}$. The choice of logical Pauli operator gives $\hat{Z}_L = \hat{Z}^{\otimes 3}$. Parallel single-qubit rotations on all qubits are followed by a round of quantum error detection, which excludes the wavefunction components with single-qubit or two-qubit flips. Thus, the final state is $\ket{+_L}-i\epsilon^3\ket{-_L}$. The inset describes the error correction/detection and the logical space stabilized by $S = \{\hat{X}_1\hat{X}_2, \hat{X}_2\hat{X}_3\}$.}\label{fig:phaseFlipCode} 
\end{figure*}

\section{Non-Clifford state preparation}
\label{sec:framework}

Figure \ref{fig:phaseFlipCode} illustrates our proposal for FT preparation of non-Clifford states with the three-qubit phase-flip code. Our protocol starts with a fault-tolerantly initialized Clifford state. For simplicity, we choose in the example $\ket{+_L}=\ket{+++}$ (Fig.~\ref{fig:phaseFlipCode}, left) \footnote{This differs from the conventional definition where $\ket{0_L} = \ket{+++}$}. We apply physical $\hat{Z}$-rotation gates, with a small rotation angle $2\epsilon$, to the three physical qubits. This results in the state $\approx\ket{+++}-i\epsilon^3\ket{---}+\text{(other terms)}$. These other terms omitted here are a linear combination of single-qubit or two-qubit flipped states, with probability amplitudes of $i\epsilon$ and $-\epsilon^2$, respectively (Fig.~\ref{fig:phaseFlipCode}, middle). Then, quantum error detection rejects the wavefunction outside the logical space, projecting the state into $\ket{+_L}-i\epsilon^3\ket{-_L}$. The resulting state is a rotated $\ket{+_L}$ state around $z$ axis by $-2\epsilon^3$. The inset of Fig.~\ref{fig:phaseFlipCode} shows the quantum error detection with the phase-flip code. If the syndrome measurements are successful, i.e. $\hat{X}_1\hat{X}_2=1$ and $\hat{X}_2\hat{X}_3=1$, the code accepts the state. Otherwise, the state is given up, and the procedure is repeated.

Our method can be generalized to any quantum error-correcting codes. Here, we first focus on the $[[n,1,d]]$ code with a logical $Z$ operator, \begin{align}
\hat{Z}_L = \displaystyle{\prod_{i=1,..,d}}\hat{Z}_i,
\end{align}
where $\hat{Z}_i$ is a Pauli-$\hat{Z}$ operator acting on the $i$-labeled qubit. Rotating all physical qubits in $\hat{Z}_L$, i.e., the support of $\hat{Z}_L$, around the $Z$ axis by $\theta$ gives
\begin{align}
    &\bigotimes_{i=1,\dotsc,d}\hat{R}_{z,i}(\theta) = \\
    &\sum_{n_I} \cos^{(d-n_I)}\left(\frac{\theta}{2}\right)\left(i\sin\left(\frac{\theta}{2}\right)\right)^{n_I} \displaystyle{\prod_{\substack{i=1,..,d \\ \#\hat{I}_i = n_I}}}\left\{\hat{I}_i,\hat{Z}_i\right\}.
\label{eq:physical_rotations}
\end{align}
In a noise-free setting, correct syndrome measurements project the qubits in the logical codeword space. Formally, the projection operator is given by $\hat{\Pi} = \ket{0_L}\bra{0_L}+\ket{1_L}\bra{1_L})$. All terms in Eq.~(\ref{eq:physical_rotations}) except $\hat{I}_L=\hat{I}_1\hat{I}_2\cdot\cdot\cdot\hat{I}_d$ and $\hat{Z}_L=\hat{Z}_1\hat{Z}_2\cdot\cdot\cdot\hat{Z}_d$ are discarded because they correspond to $1$ to $d-1$ qubit phase-flip errors. We have 
\begin{align}
&\hat{\Pi}\cdot\left[\bigotimes_{i=1,\dotsc,d}\hat{R}_{z,i}^\dagger(\theta)\right] = \\
&\frac{1}{\sqrt{\cos^{2d}(\theta/2)+\sin^{2d}(\theta/2)}}(\cos^d(\theta/2)\hat{I}_L+i\sin^d(\theta/2)\hat{Z}_L),
\label{eq:logical_rotation}
\end{align}
where, without loss of generality, we assume $i^d = i$. (See Appendix~\ref{app:non-rotation} for when the assumption is relaxed). Therefore, the above procedure accomplishes a logical $Z$-axis rotation by the angle: 
\begin{align}
\theta_L = 2\sin^{-1}\left(\frac{\sin^d(\theta/2)}{\sqrt{\cos^{2d}(\theta/2)+\sin^{2d}(\theta/2)}}\right). \label{eq:logical_angle}
\end{align}

As such, our protocol implements a logical rotation via physical rotations and quantum error detection. We denote,
\begin{align}
    \bigotimes_{i=1,\dotsc,d}\hat{R}_{z,i}^\dagger(\theta) \xrightarrow{\text{QED}} \hat{R}_{z,L}^\dagger(\theta_L),
\end{align}
where ``QED" stands for the quantum error detection and post-selection process.  Furthermore, this framework can be used directly to prepare magic states of arbitrary angle $\theta_L$:
\begin{align}
    \ket{M_{\theta_L}} = \hat{R}_{z,L}^\dagger(\theta_L) \ket{+_L}. \label{eq:magic_state}
\end{align}

The framework is generalized to any quantum stabilizer codes. For a non-CSS code, $\hat{Z}_L = \bigotimes_{i=1,\dotsc,d}\hat{P}_i$ where $\hat{P}_i$ is a Pauli operator of the $i^\text{th}$ qubit. We can achieve arbitrary-angle, logical rotation $\hat{R}_{z,L}(\theta_L)$ by applying physical rotations on each qubit on the corresponding Pauli basis,
$\hat{R}_{z,L}(\theta_L) = \bigotimes_{i=1,\dotsc,d}\hat{R}_{P_i}(\theta)$ followed by error detection. Alternatively, one can use the inverse Pauli frame mapping. Denoting a physical Pauli operator with a prime($'$), it is possible to express the normalizer $\bigotimes_{i=1,\dotsc,d}\hat{P}_i=\bigotimes_{i=1,\dotsc,d}\hat{Z}_i'$. Thus, we can use the same procedure with $\hat{Z}_i'$ rotations as far as the stabilizers are also defined in the mapped frame. The benefit of the latter approach is that we can fully use precise virtual gates. 

We can apply our method to $[[n,k\neq1,d]]$ codes. We discuss in Appendix~\ref{app:4qubit} the example of the 4-qubit error detection code.

\subsection{Error Rates}

In this section, we analyze our protocol's error suppression capability. To do so, we consider incoherent and coherent error rates of the resulting states separately.

\subsubsection{Incoherent Errors}
Incoherent errors are modeled as stochastic applications of the Pauli gates. Here, we use single-qubit Pauli errors $\hat{X}, \hat{Y}, \hat{Z}$, each with probability $p_{\text{in}}/3$. $\hat{X}$ and $\hat{Y}$ errors are detected and rejected independently of the logical rotations. However, a physical $\hat{Z}$ error can pass error detection if it simultaneously happens with the physical rotations.

Consider that one of the physical qubits in $\hat{Z}_L$ experiences a single Pauli-$\hat{Z}$ error. Normally, this is detected by the substrate code. However, if the error occurred during the logical rotation operation (Eq.~(\ref{eq:logical_rotation})), it could pass through the detection with nonzero probability. The leading-order error is a single-qubit $\hat{Z}$ error on any of the $d$ qubits along the logical operator. This error is seldom undetected in the projection to the logical space. Consider again the example of a 3-qubit phase-flip code in Fig.~\ref{fig:phaseFlipCode}. If we apply the phase flip of the first qubit on the state of Eq.~(\ref{eq:magic_state}),
\begin{align}\label{eq:phaseFlipCodeError}
    \ket{-++}+\epsilon^3\ket{+--} + \epsilon(\ket{+++}+\epsilon\ket{---}) + (...) .
\end{align}
Thus, with the probability $\approx p_\text{in}/3\cdot \epsilon^2$, the error is undetected resulting in the state $\ket{+++}+\epsilon\ket{---}$.

We generalize the errors resulting from single-qubit errors. The probability of the error is
\begin{align}
    p^L_\text{in} = d'\frac{p_{\text{in}}}{3}\cos^{2(d-1)}(\theta/2)\sin^2(\theta/2),
\end{align} 
where $d'$ is the number of possible single-qubit errors that are possibly undetectable. However, the error state is not orthogonal to the desired state, and the error detection protocol discards the state with a finite probability. Thus, for a fair comparison with the usual incoherent logical error rate, we define the effective error rate of the post-selected state.
\begin{align}
    \epsilon_\text{in} &= p^L_\text{in}\cdot\left|\braket{\psi_e|\psi}\right|^2 \cdot p_\text{s}^{-1} \nonumber \\
    & \approx d'\frac{p_{\text{in}}}{3}\sin^{2(d-1)}(\theta/2)\cos^{-1}(\theta/2), \label{eq:firstOrderError}
\end{align}
where $\ket{\psi_e}$ is the error state, and $p_\text{s}$ is the success probability of post-selection.

More generally, $n$ errors can occur during the logical rotation, and the total error rate can be written as 
\begin{align} 
\epsilon_\text{in} \approx \sum_{n}d'(d,n,\mathscr{C}) \left(\frac{p_\text{in}}{3}\right)^n\sin^{2(d-n)}(\theta/2)\cos^{-2n}(\theta/2), \label{eq:inc_error}
\end{align}
where $d'$ is a function of distance $d$, number of errors $n$, and code $\mathscr{C}$. $\epsilon_\text{in}$ is the dominant error of our protocol in the low physical error regime $p_\text{in}<\sin^2(\theta/2)$. If $p_\text{in}>\sin^2(\theta/2)$, the error rate is limited by the logical error rate of the substrate code, e.g., $p_L \sim C(p/p_\text{th})^d$ with constant $C$ and the error threshold $p_\text{th}$. Furthermore, in a low physical error regime, the first order error of Eq.~(\ref{eq:firstOrderError}) dominates the total error. 

Note that the error in Eq.~(\ref{eq:firstOrderError}) is exponentially suppressed with the code distance: $\epsilon_{in} \sim \mathcal{O}(\sin^{2(d-1)}(\theta/2))$. Moreover, we call this property \emph{continuous fault-tolerance}, as the rate of the exponential suppression is determined by an adjustable parameter $\theta$. 


Incoherent errors of physical qubits can induce logical coherent errors, i.e., with rotation angle $\approx(\theta/2)^{d-2}$. It is an open question how the coherence of errors in individual gates affects the fidelity of the algorithm. Here, we assume that an algorithm uses many rotation gates on different axes so that the coherence of individual rotation does not add up. This justifies the use of an effective error rate in Eq.~(\ref{eq:inc_error}). 

\begin{figure}
    \centering
    \includegraphics[width=0.8\columnwidth]{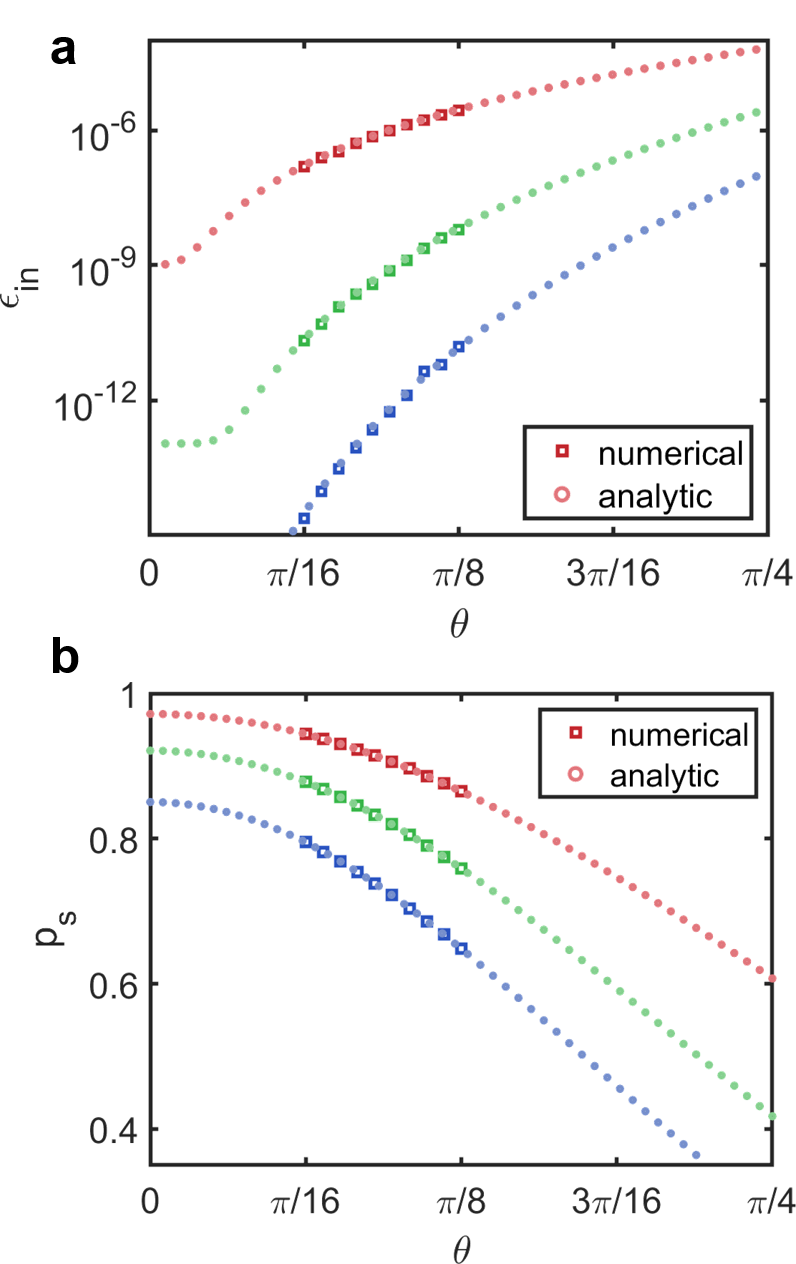}
    \caption{\textbf{Non-Clifford state preparation with rotated surface codes.} \textbf{a} Physical rotation angle ($\theta$) vs. effective error rate ($\epsilon_\text{in}$) of the resulting state and \textbf{b} physical rotation angle  vs. success rate, for code distance $d = 3, 5, 7$ in red, green and blue, respectively. The analytic expression with only first-order errors (Eq.~\ref{eq:firstOrderError}, circle) matches well with the numerical Clifford simulation results (square). See Appendix~\ref{app:sim} for more details. For all code distances, error detection is applied twice to suppress the readout errors (see Appendix~\ref{app:surfaceCodeErrorTypes}). Input physical error rate is set to $p_{in} = 10^{-3}$.}\label{fig:SimResult}
\end{figure}

\subsubsection{Coherent Errors}

Physical coherent errors could be more detrimental to our protocol because it is fundamentally indistinguishable from our logical rotation operation. Fortunately, $\hat{Z}$ rotation of a physical qubit is equivalent to adjusting the phase of local oscillators driving the qubit. In this case, the $\hat{Z}$ rotation has only angle errors but no axis errors. Moreover, these so-called virtual gates are very precise by locking with the atomic clock, which achieved a fractional error less than $10^{-18}$~\cite{brewer2019al+}.

We analyze coherent errors for current and near-term devices in an extra-conservative evaluation fashion. We consider the error model where the rotation angle of $i$-labeled physical qubit $\theta_i$ deviates from the target angle $\theta$ by $\Delta\theta_i$ ($\theta_i = \theta + \Delta\theta_i$). Here $\Delta\theta_i$s are assumed to be independently and identically distributed with Gaussian standard deviation $\sigma_\theta$. That is, $\Delta\theta_i \sim \mathcal{N}(0, \sigma^2_\theta)$.

Assuming $\theta_i = \theta \ll 1$ and the fractional error $\Delta\theta_i/\theta_i$ are small, $\theta_L\approx2\left(\frac{\theta}{2}\right)^d(1+2\sum_{i}\frac{\Delta\theta_i}{\theta})$. By the central limit theorem (CLT), the fractional logical rotation angle deviation, $\delta$, from the target angle ($\theta_{L,0} \approx 2(\theta/2)^d$) follows a normal distribution. That is, the angle error is
\begin{align}
\Delta\theta_L \sim \mathcal{N}\left(0,\left(\sqrt{d}\theta_{L,0}\frac{\sigma_\theta}{\theta}\right)^2\right).
\end{align} 
Thus, the error in the logical rotation angle scales with the square root of the code distance $d$. 

Indeed, coherent rotation errors are often neglected in distillation literature because magic states can be twirled, averaging out coherence~\cite{bravyi2012magic,duclos2015reducing,o2017quantum}. In addition, repetitive application of rotation would average its angle error by the CLT as our method. However, these arguments are optimistic scenarios, while the error will highly depend on the application and the size of the system.

\subsection{Success Rate}\label{sec:discard}
In error detection, we trade off resource states for lower error rates. We should discard the states with $\hat{X}$-stabilizer flip ($\hat{Z}$ error). This prevents the states in a more rotated state, e.g., $\ket{-++}+\epsilon\ket{+--}$ state in our phase-flip code example. We also need to discard the states with $\hat{Z}$-stabilizer flip because we do not have prior knowledge of whether this is due to correctable $\hat{X}$ error or the combination of uncorrectable $\hat{Z}$ error and correctable $\hat{Y}$ error. Similarly, $\hat{Y}$ errors with both stabilizers flipped are rejected.

A logical state is successfully prepared if no error (either from an incoherent source or a coherent source) has been detected. Thus, we analytically derive the success rate,
\begin{align}
    p_s &\approx p_s^\text{in}p_s^\text{coh}, \label{eq:discard_rate} \\
    p_s^\text{in} &\approx (1-p_\text{in})^{r\cdot n}(1-2p_\text{in}/3)^{r\cdot(n-1)},\label{eq:succ_rate_in}\\
    p_s^\text{coh} &\approx \cos^{2d}(\theta/2)+\sin^{2d}(\theta/2). \label{eq:succ_rate_coh}
\end{align}
where $p_{\text{in}}$ is the input physical error rate and $r$ is the number of stabilizer measurements to suppress the measurement error, and $p_s^\text{in}$ ($p_s^\text{coh}$) is the success rate after discarding by the detection of incoherent (coherent) errors. 

Fig.~\ref{fig:SimResult}\textbf{b} shows the success rate of the state preparation on a surface code with the distances 3 (red), 5 (green), and 7 (blue). As shown, the analytical results (square) by Eq.~(\ref{eq:discard_rate}) agree with the numerically simulated results (circle). 

The proposed state preparation framework works particularly well (in terms of both error rate and success rate) for small angles, as shown in Fig.~\ref{fig:SimResult}. The next section introduces novel schemes for implementing fault-tolerant arbitrary-angle rotation gates (from small-angle magic states) with low overhead.

\section{Fault-Tolerant Gate Schemes}

\begin{figure*}
    \centering
    \includegraphics[width=0.9\textwidth]{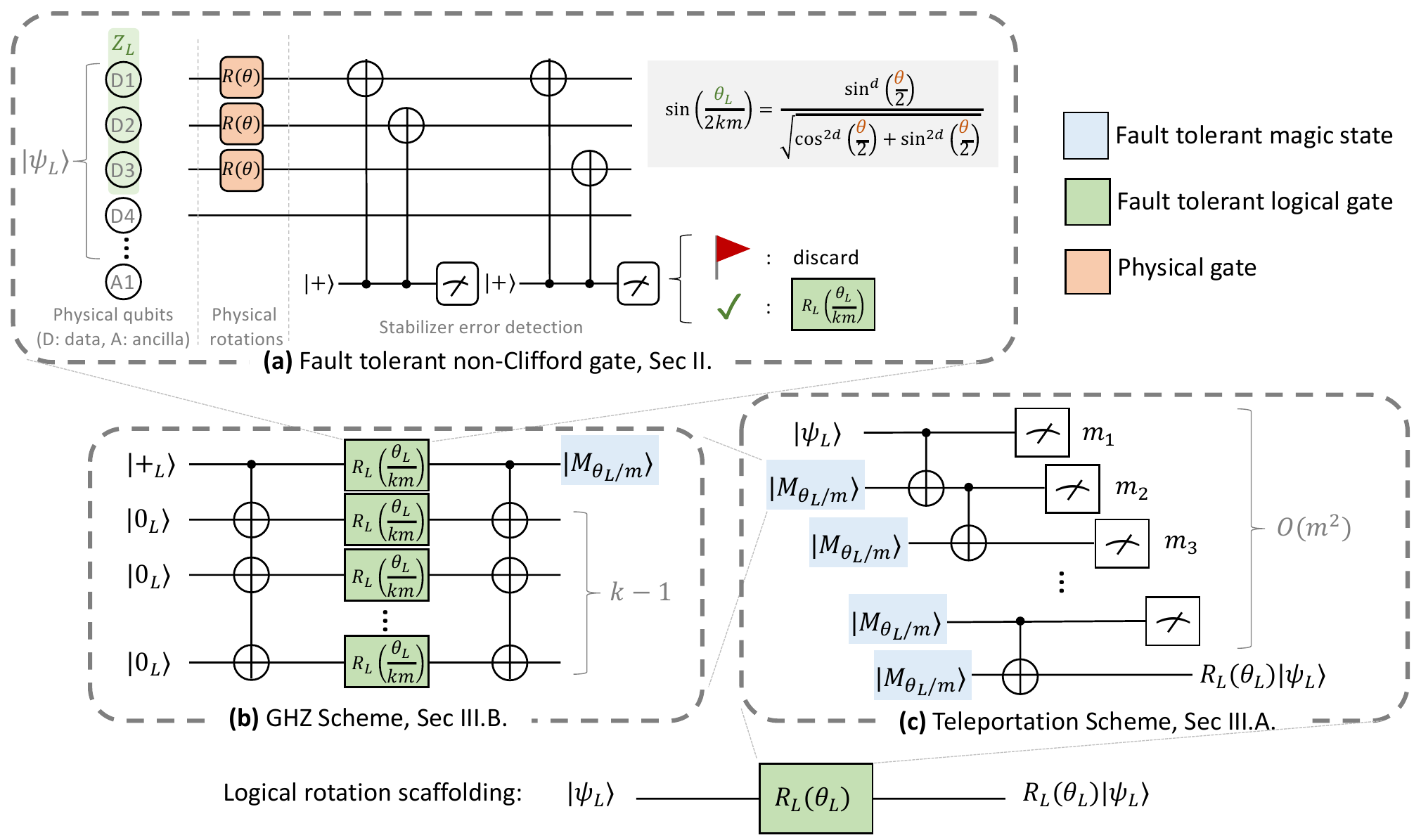}
    \caption{\textbf{Fault-tolerant gate schemes.} (a) Example implementation of the proposed fault-tolerant non-Clifford gate method from Section~\ref{sec:framework}. A logical rotation is accomplished by physical rotations followed by error detection. (b) GHZ scheme for synthesizing rotation gates from Section~\ref{sec:ghz}. In this example, $k$ logical rotations, $R_L(\theta_L/km)$, are performed in parallel, resulting in an overall rotation by the angle $\theta_L/m$. (c) Teleportation scheme for synthesizing rotation gates from Section~\ref{sec:teleportation}. The example circuit teleports an expected $O(m^2)$ small-angle magic states, $\ket{M_{\theta_L/m}}$, to achieve a $R_L(\theta_L)$ gate. The exact rotation is determined by the measurement outcome $m_i$ in each step, i.e., the resulting angle $\theta_L' = \sum_i(-1)^{m_i}\theta_L/m$. Overall, the rotation scaffolding technique in Section~\ref{sec:scaffolding} optimizes the parameters $k,m$ to implement $\theta_L$ with the lowest overhead and highest fidelity possible.}
    \label{fig:FTGate}
\end{figure*}

This section introduces how to use the proposed state preparation framework to implement fault-tolerant non-Clifford gates, such as arbitrary-angle rotation gates. 
The conventional wisdom for fault-tolerantly implementing a non-Clifford gate $U$ is to first synthesize $U$ as a sequence of Clifford and $T$ gates and then implement $T$ gates by magic state distillation. However, such implementation incurs significant overhead, as the synthesized sequence can be long and the cost for $T$ magic state distillation is high. 

Given the proposed fault-tolerant state preparation framework, we are now ready to discuss how to efficiently implement arbitrary-angle rotation gates. A naive approach is to directly rotate the logical qubits using Eq.~(\ref{eq:logical_rotation}) by $\theta_L$ or consecutively rotate $m$ times by $\theta_L/m$. However, as shown in Section~\ref{sec:discard}, the limited success probability inhibits scaling up to large angles $\theta_L$. 

To overcome this, we demonstrate two improved schemes with substantially lower overhead -- one is \emph{sequential} based on gate teleportation and the other is \emph{parallel} based on GHZ states. We also discuss their practical implementations and trade-offs in terms of resource overhead. Finally, we introduce a technique called ``rotation scaffolding'' to balance the space-time cost, by hybridizing the teleportation and GHZ schemes.

\subsection{Gate Teleportation Scheme}\label{sec:teleportation}


Fig.~\ref{fig:FTGate}\textbf{c} shows the teleportation scheme for synthesizing a rotation gate, $R_{\theta_L}$, with a target angle $\theta_L$. Consider using the magic states $\ket{M_\theta}$ that are prepared by Eq.~(\ref{eq:magic_state}), for some small angle $\theta$. For simplicity, we assume $\theta_L = m\theta$ for some integer $m$. Starting with the initial $\ket{+}$ state, we teleport magic states $\ket{M_\theta}$ consecutively until the output state is either $R_{\theta_L}\ket{+}$ or $R_{\theta_L}^\dagger\ket{+}$. Let the number of teleportation steps be $\chi$. For example, in Fig.~\ref{fig:FTGate}\textbf{c}, $\chi = 3$, that is, three magic states are teleported. The resulting state is rotated by the angle $\theta' = \sum_i (-1)^{m_i}\theta$,  where $m_i$ are the measurement outcomes of the circuit. In particular, the $\chi$ measurements correspond to $\chi$ fair coin tosses. If a measurement has outcome ``0'', the state is rotated by $\theta$ and if the outcome is ``1'', the state is rotated by $-\theta$. Therefore, the process follows a \emph{one-dimensional random walk}. The number of ``0'' measurement outcomes are normally distributed with mean $\chi/2$ and standard deviation $\sqrt{\chi}/2$. Therefore, to reach $\theta_L$ or $-\theta_L$, we need the number of the measurements outcomes ``0'' and ``1'' to differ by $m$. Thus, we need on average $\bar{\chi} \sim O(m^2)$ trials. Subsequently, if $R_{\theta_L}^\dagger\ket{+}$ is reached, one can recover $R_{\theta_L}\ket{+}$ by a simple Pauli-X correction. Overall, $O(m^2)$ magic states are teleported sequentially. In the case that all magic states are prepared in advance, we would need $O(m^2)$ space cost to store them; however, in our scheme, the time it takes to produce $\ket{M_\theta}$ is on the same order of the teleportation time, so we can reuse the same qubits for preparing and teleporting the magic state, reducing the space cost to $O(1)$. As such, the teleportation scheme consumes $O(m^2)$ space-time cost.


\subsection{Greenberger–Horne–Zeilinger (GHZ) Scheme}\label{sec:ghz}

Fig.\ref{fig:FTGate}\textbf{b} shows a scheme constructed from the GHZ state. To synthesize a fault-tolerant rotation gate $R_{\theta_L}$, we use logical rotations $R_\theta$ from Eq.~(\ref{eq:logical_rotation}), where $\theta = \theta_L/m$. In this scheme, we apply a multitarget CNOT gate to create an $(m+1)$-qubit GHZ state. Subsequently, $m$ parallel logical rotations are performed, each with a success probability $p_s$ (Eq.~(\ref{eq:discard_rate})). Finally, another round of the multitarget CNOT gate disentangles the qubits to accomplish a combined rotation, i.e., $R_{m\theta}\ket{+}$. The multi-target CNOT gate can be implemented efficiently and fault-tolerantly with lattice surgery or braiding in surface code. Furthermore, the logical rotations are performed in parallel and errors are suppressed via error detection. The number of qubits used in this scheme is linear, that is, $O(m)$ space cost. However, if one of the parallel trials fails, all logical qubits need to be discarded and repeated. Fortunately, for a fixed angle $\theta_L$ and practical range of $m$, as $m$ increases, the success rate of $m$ parallel $R_{\theta_L/m}$ rotations \emph{increases}, thanks to the properties that $\left(p_{s}^{coh}\right)^m \to 1$ and $p_{s}^{in}$ is independent of $m$. 
The average number of steps is approximately $\chi = p_s^{-m}$, which results in the time cost $O(1)$ for $m\lesssim \log_{10}p_s$ thanks to parallelism \footnote{In practice, the time cost for multi-target CNOT gate depends on parameters such as the code distance. A more rigorous resource overhead analysis is presented in later sections.}. Therefore, for small angles $\theta$ (when $p_{s}$ is large), such a parallel scheme can be advantageous in overall space-time cost.


\subsection{Rotation Scaffolding: A Hybrid Scheme}\label{sec:scaffolding}

We now describe a hybrid method that combines the best of each scheme. We name this technique \emph{rotation scaffolding}, for it implements rotation gates fault tolerantly by combining small-angle logical rotations and magic states strategically, as illustrated in Fig~\ref{fig:FTGate}. Let $k, m \in \mathbb{N}$ be the parameters of the scheme to be optimized. Suppose that we want to implement an arbitrary-angle logical rotation $R_L({\theta_L})$, using the smaller-angle logical rotations $R_L(\theta_L/mk)$ from Eq.~(\ref{eq:logical_rotation}). Because the GHZ scheme works well for small rotation angles, we will use it to boot-strap into magic states $\ket{M_{\theta_L/m}}$. Finally, we use multiple copies of magic states $\ket{M_{\theta_L/m}}$ in the teleportation scheme to produce the desired $R_L({\theta_L})$.

As we showed in the previous schemes, the space-time cost of preparing $\ket{M_{k\theta}}$ scales as $O(k)$ for small $k$ and small angle $\theta$ and $O(\exp(k))$ for large $k$. We then use the $O(m^2)$ copies of the magic states to produce $\ket{M_{\theta_L}}$. In principle, we obtain an overall scaling of $O(m^2k)$ in space-time cost, assuming we started with small-angle rotations $R_L(\theta)$ and appropriate choices for $m,k$. To find the optimal parameters, we iterate over the possible combinations and choose the one with minimal space-time overhead. Details on the implementation of this scheme are given in Section~\ref{sec:resource}.

\subsection{Resource Overhead Comparison}\label{sec:resource}

\begin{figure*}
    \centering
    \includegraphics[width=\textwidth]{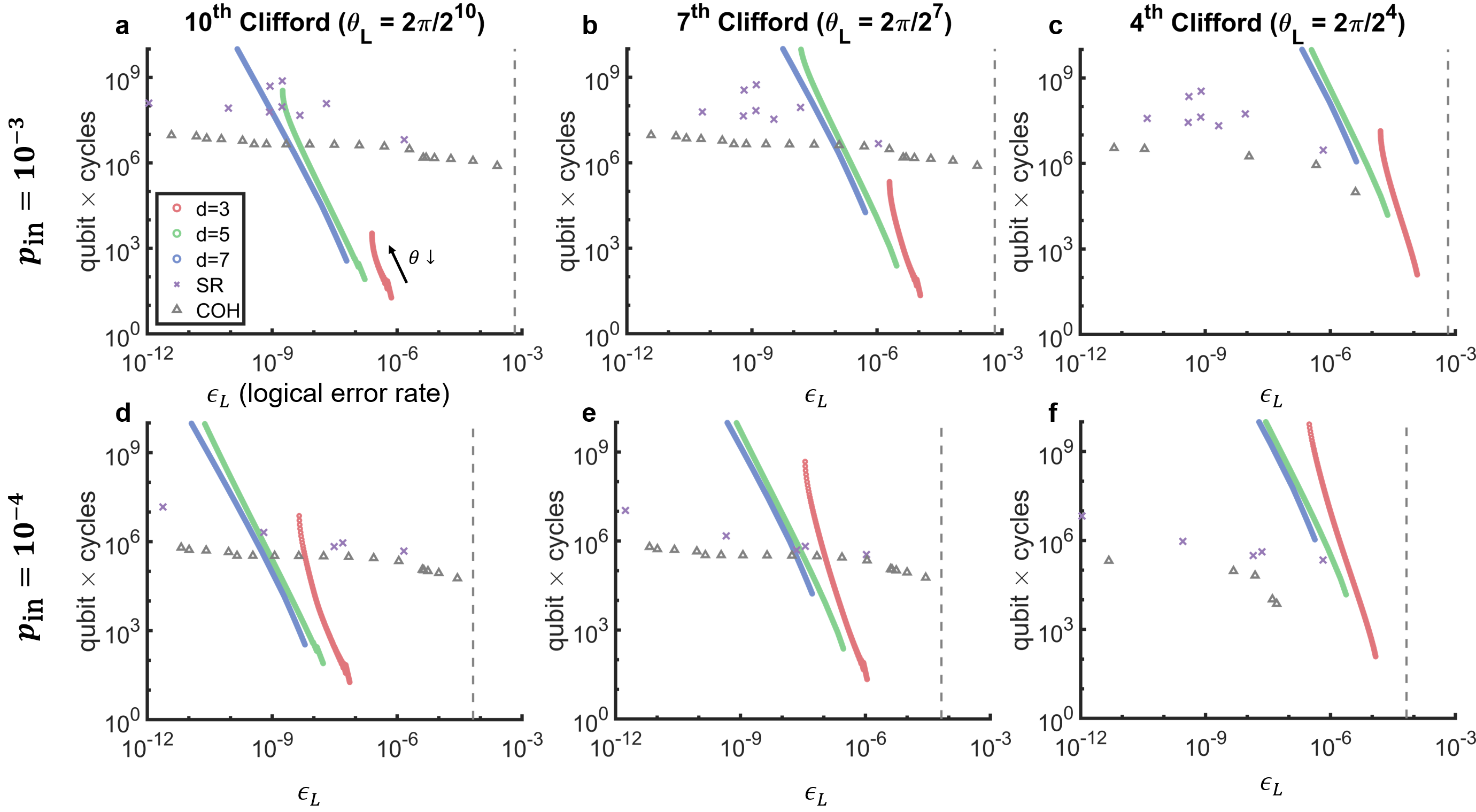}
    \caption{\textbf{Resource overhead (vertical axis) of preparing higher-order Clifford states on the surface code substrate as a function of the resulting logical error rate (horizontal axis).} The space-time costs for preparing magic states in the 10$^\text{th}$ (left), 7$^\text{th}$ (middle), and 4$^\text{th}$ (right) Clifford hierarchy are shown. Upper (lower) panels assume a single-qubit error rate of $p_\text{in} = 10^{-3} (10^{-4})$. $\circ$: Our method to prepare $\ket{\theta_L}$ from different physical rotation angles $\theta$, with distances 3 (red), 5 (green), and 7 (blue). Decreasing angle $\theta$ reduces the logical error rate $\epsilon_L$ while the cost increases. When $\theta$ is below some angle, decreasing $\theta$ does not help, i.e., increasing both cost and $\epsilon_L$. These sub optimal data points are not shown in the plot. Clifford overheads and errors in angle combinations associated with it are ignored (See Appendix~\ref{app:Clifford} for the Clifford cost). $\times$: Selinger-Ross approximated rotation synthesis with pre-distilled $T$ states~\cite{ross2014optimal}. In this calculation, we ignored the Clifford overheads and Litinki's table~\cite{litinski2019magic} was used for the distillation cost. $\triangle$: Campbell-O'Gorman-Howard parity checker method~\cite{campbell2016efficient,campbell2018magic}. Likewise, Clifford overhead is ignored, and the cost mainly derives from $T$-distillation \cite{litinski2019magic}. The Clifford overhead of the COH method is non-negligible, so the estimation sets the lower bound. 
    }
    \label{fig:resource}
\end{figure*}

We compare the resource overhead of our method with that of the existing methods using distillation and gate synthesis. Here, we define the resource overhead as the space-time cost, the number of qubits multiplied by the number of time steps that are occupied for computation. A time step is defined by a full stabilizer measurement cycle. We consider the surface code as a substrate, as in previous sections. Because the resource overhead depends on the final error rate, we compare them in a 2D parameter space of the resource overhead and the logical error rate.

For the distillation, we used the results of \cite{litinski2019magic} with surface code that includes the cost of Clifford gates and qubit routing. In the work, Litinski also combined multi-qubit rotation circuits and faulty $T$ measurements on the surface code and optimized asymmetric code distances for $\hat{X}$, $\hat{Z}$, and measurement errors \cite{litinski2019magic}. The implementation achieved orders of magnitude improvement in the space-time cost compared to the previous state-of-the-art. We synthesize 10$^\text{th}$, 7$^\text{th}$, and 4$^\text{th}$ Clifford states with $T$-equivalent ancilla states using two methods. 

The conventional way to synthesize a rotation gate is using $T$ magic states and Hadamard ($H$) gates proposed by Ross and Selinger~\cite{ross2014optimal}, which we call the RS method. The RS method synthesizes single-qubit $z$ rotations with an arbitrary angle. The resulting rotation is approximate, and there exists the errors in rotation angle, $\epsilon$. This rotation angle error is a coherent error, and a one-to-one comparison of the method with ours is not apple-to-apple. For benchmarking purposes, we present only the incoherent error of the RS method for $\theta_\epsilon\approx\theta_L/10$ starting from the distilled $T$ states. The number of $T$ magic states for synthesizing the state with angle error $\theta_\epsilon$ is $n_T \approx 3\log_2(1/\theta_\epsilon) + \mathcal{O}\left(\log(\log(1/\theta_\epsilon))\right)$ that saturates the information-theoretic bound. 

In the second method proposed by Campbell and O'Gorman~\cite{campbell2016efficient} and subsequently improved by Campbell and Howard~\cite{campbell2018magic}, which we call the COH method, one synthesizes a fault-tolerant $l+1^\text{th}$ Clifford state using two $l+1^\text{th}$ Clifford states, one $l^\text{th}$ Clifford state, and 8 $T$ magic states with errors $\epsilon_{l+1}$, $\epsilon_l$ and $\epsilon_{T}$ respectively. The circuit measures the parity of two $l+1^\text{th}$ Clifford states to suppress the errors of the resulting $l+1^\text{th}$ states to $\epsilon'_{l+1}\approx 8\epsilon_T^2+\epsilon_{l+1}^2+\frac{1}{4}\epsilon_l$. The synthesis proceeds from $T$ magic state to distill $4^\text{th}$ Clifford state, which is used for $5^\text{th}$ state generation, and so on. 

Figure~\ref{fig:resource}\textbf{a} shows the result of the resource overhead and error calculation for the resource state for 10$^\text{th}$-order Clifford gate, i.e., $R_z(2\pi/2^{10})\ket{+}$. For our method ($\circ$), we consider the rotation scaffoliding starting from different small rotation angles and choose the lowest cost for a given error probability. For RS ($\times$) and COH methods ($\triangle$), we started from the errors and space-time costs of $T$-magic states by distillation with a rotated surface code lattice~\cite{litinski2019magic} and calculated the resulting errors and costs from the number of $T$-magic states. In this calculation, we neglected the overhead from Clifford gates and space-time layout, and see Appedix~\ref{app:Clifford} for the discussion. As shown in Fig.~\ref{fig:resource}, our method is especially advantageous in the $\epsilon_L \approx 10^{-8}\sim 10^{-4}$ regimes. For a lower error, COH is effective due to a quadratic error suppression in parity checking. We emphasize that our method can be combined with COH method by first generating relatively high error rate states with our method and distilling it with the COH. Due to the low error raw states, this combination can offer a very fast reduction of error, with a small number of pre-distilled $T$ states. Note that our method has a preserved (but narrower) advantage for a lower physical error ($p_\text{in}$) and lower order Clifford states (Fig.~\ref{fig:resource}\textbf{f}).

To summarize, our method achieves small angle rotation with ultralow error and negligible cost. Moreover, for a larger angle, the method fills the gap of the intermediate error regime between NISQ and fault tolerance. 

\section{Conclusion}

We have introduced a novel fault-tolerant scheme for logical non-Clifford single-qubit rotation gates. Our implementation of logical gates is based on preparing arbitrary-angle magic states through error detection. With today's hardware error rates ($10^{-4}$ to $10^{-3}$) and a small error detection code, we can implement arbitrary-angle rotation gates with a logical error rate of $10^{-8}$ to $10^{-4}$ using a factor of $10^2$ to $10^4$ lower qubit$\times$cycle. 

With these building blocks, our protocol is immediately useful for boostrapping NISQ hardware into near-term fault-tolerant systems. Quantum algorithms -- especially those with many non-Clifford gates, such as small-angle rotations -- would require orders-of-magnitude lower resources and thus be executed sooner than otherwise possible. Conventionally, a logical small-angle rotation is constructed using many magic $T$-gates~\cite{campbell2017unifying}. In our method, small-angle rotations can be implemented directly with ultra-high fidelity after error detection.
Furthermore, our rotation scaffolding protocol uses multiple small-angle rotations to construct a larger-angle rotation with the lowest overhead and highest fidelity possible. Through a systematic comparison, we have shown that the advantage of our protocol over distillation or synthesis methods depends on what rotation angles are needed in a circuit, what physical error rate we have, and what target logical rate is needed.   

Our implementation of logical rotation through post-selection has a few additional advantages: 1) it suppresses physical errors via quantum error detection, implying it achieves full distance-$d$ protection compared to a $\lfloor d/2 \rfloor$ protection by error correction of the same code; 2) furthermore, our method can also be used as state injection for parity checker protocols; for example, the resource state generated by our method can be used with COH method to suppress remaining errors; 3) the protocol, specifically success rate, can be improved by carefully treating non-diagonal errors that are not affected by the physical rotations. Future researchers can find the optimal way of the hybrids, detecting and correcting errors. It is also an interesting direction to explore our proposal for the use of an error-detectable gate, e.g. in low-depth circuits.

Our method is also compatible with newly developed techniques such as bias-preserving error correction code \cite{puri2020bias}, or noise-balanced XZZX-surface codes \cite{bonilla2021xzzx}. Moreover, a direct distillation of small angle rotation generated from our method is plausible to extend the benefiting regime. We leave benchmarking our method under various codes and new distillation methods including spacetime cost for routing for future research. 

\section{data availability}
All the numerical data presented in this paper are the results of Julia and Matlab simulations. The code used to generate this data is open-sourced, available in a GitHub repository~\cite{yongshan2022SmallAngleGit}.

\section{acknowledgement}
We thank Isaac Chuang, Steve Girvin, Jonathan Baker, and Wenxuan Xie for the fruitful discussion. This work was partially supported by MITRE, NSF CQN, NSF EPiQC, and Yale cluster. H.C acknowledges the Claude E. Shannon Fellowship and the Samsung Scholarship.


%

\appendix

\section{Non-rotation operators}\label{app:non-rotation}

In this appendix, we discuss the even distance ($d$) code under physical rotations and post-selection. For even $d$, the amplitudes of products of operators, $\hat{I}^{\otimes d}$ and $\hat{Z}^{\otimes d}$  
Following the framework of main text (Eq.~(\ref{eq:physical_rotations})), the operation after post-selection is
\begin{align}
    \hat{O} = \cos^d(\theta/2)\hat{I}_L - (-1)^{d/2} \sin^d(\theta/2)\hat{Z}_L.
\end{align}
We assume $(-1)^{d/2} = -1$ without loss of generality.

Apart from normalization, $\hat{O}$ is not unitary and cannot be a (logical) single-qubit rotation gate. In fact, the action of $\hat{O}$ depends on the state on which it acts. For example, $\hat{O}$ in the state $\ket{0_L}$ is $\hat{O}\ket{0_L} = (\cos^d(\theta/2)-\sin^d(\theta/2)) \ket{0_L}$. On the other hand, $\hat{O}\ket{1_L} = (\cos^d(\theta/2)+\sin^d(\theta/2)) \ket{1_L}$. For general quantum state, the action of $\hat{O}$ reduces the $\ket{0_L}$ component while increasing $\ket{1_L}$ component. This ``(weak) quantum state filtering'' can be a new toolkit in quantum algorithm designs. 

We also note that for the states that are not $\ket{0_L}$ and $\ket{1_L}$, $\hat{O}$ is approximately the rotation around the equatorial axis perpendicular to the great circle on Bloch sphere where the state is. Thus, it can serve as a rotation about a state-dependent axis by a state-dependent angle.

\section{Incoherent error for multi-rotation by small angles}
One can perform the logical rotation multiple times in an attempt to average out, and thus reduce the error. Instead of a one-shot rotation by $\theta_{L,0}$, we consider consecutively rotating the logical qubit $m$ times by $\theta_{L,0}/m$. The physical rotation angle is adjusted to be $\theta/\sqrt[d]{m}$. 

The incoherent logical error rate is a function of the distance $d$, the physical angle $\theta$, and the number of repetitions $m$. As such, it is modified to be $\epsilon \approx md'\frac{p_{\text{in}}}{3}\sin^{2(d-1)}\left(\frac{\theta}{2\sqrt[d]{m}}\right)\cos^2\left(\frac{\theta}{2\sqrt[d]{m}}\right)$ in leading order. For small $\theta/\sqrt[d]{m}$, $\epsilon \approx d'\frac{\sqrt[d]{m^2}}{m}\frac{p_{\text{in}}}{3}\left(\frac{\theta}{2}\right)^{2(d-1)}$. Therefore, to leading order, the incoherent error rate $\epsilon \sim O(1/m^{\left(1-\frac{2}{d}\right)})$. In the large distance $d$ regime, $\epsilon$ scales as inversely as the number of repetitions, $\epsilon \sim 1/m$. Here, we see the error-time trade-off, i.e., the product of the logical error rate and the time cost is constant. In the small $d$ regime, the factor $1/m^{\left(1-\frac{2}{d}\right)}$ helps reduce the overall incoherent error rate, which is important in practical application.

\section{Coherent error for multi-rotation by small angles}

For analyzing coherent errors, we assume that the fractional error in the rotation angle is independent of rotation angles ($\Delta\theta_i$s are i.i.d with standard deviation $\frac{\sigma_\theta}{\sqrt[d]{m}}$). Each operation results in the logical rotation angle $\theta_l \sim \mathcal{N}\left(\theta_{L,0}/m,\left(\sqrt{d}\frac{\theta_{L,0}}{m}\frac{\sqrt[d]{m}}{\theta}\frac{\sigma_\theta}{\sqrt[d]{m}}\right)^2=\left(\sqrt{d}\frac{\theta_{L,0}}{m}\frac{\sigma_\theta}{\theta}\right)^2\right)$. $m$ repeated operations result in the rotation angle $\theta_L = m\theta_l \sim \mathcal{N}\left(\theta_{L,0},\left(\sqrt{d}\theta_{L,0}\frac{\sigma_\theta}{\theta}\right)^2\right)$. Thus, the fractional error distributes with $\delta \equiv (\theta_{L}-\theta_{L,0})/\theta_{L,0}\sim \mathcal{N}\left(0,\sqrt{\frac{d}{m}}\frac{\sigma_\theta}{\theta}\right)$. Note that the fractional error of the rotation angle scales as $1/\sqrt{m}$. 

\section{Example: Surface code}
The surface code is a topological error correction code. In the surface code, the stabilizer measurement and logical gate operation only require nearest-neighbor interaction in planar layout, which is suitable for large-scale monolithic fabrication of the qubits. Moreover, its high threshold of $\sim 0.75\%$ for depolarizing error keeps the error correction overhead minimal.

\begin{figure}[h]
    \includegraphics[width=0.9\columnwidth]{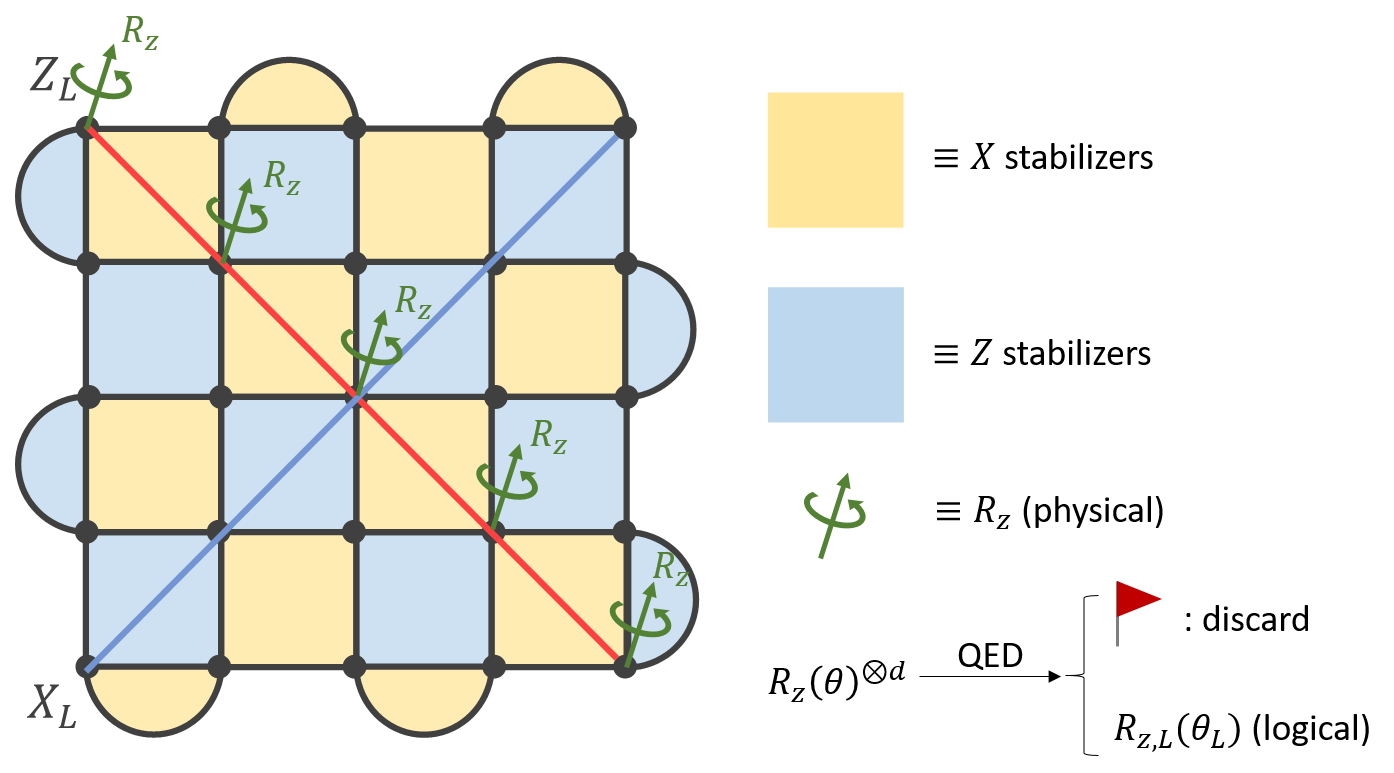}
    \caption{Rotated surface code. Yellow (blue) tiles represent the $\hat{Z}$ ($\hat{X}$) stabilizer measurement of the vertices. The logical Pauli operator, $\hat{Z}_L$ ($\hat{X}_L$) is the product of $\hat{Z}$ ($\hat{X}$) of the diagonal qubits on red (blue) lines.}
\end{figure}

\subsection{Types of logical errors} \label{app:surfaceCodeErrorTypes}

\begin{figure*}
    \centering
    \includegraphics[width=\textwidth]{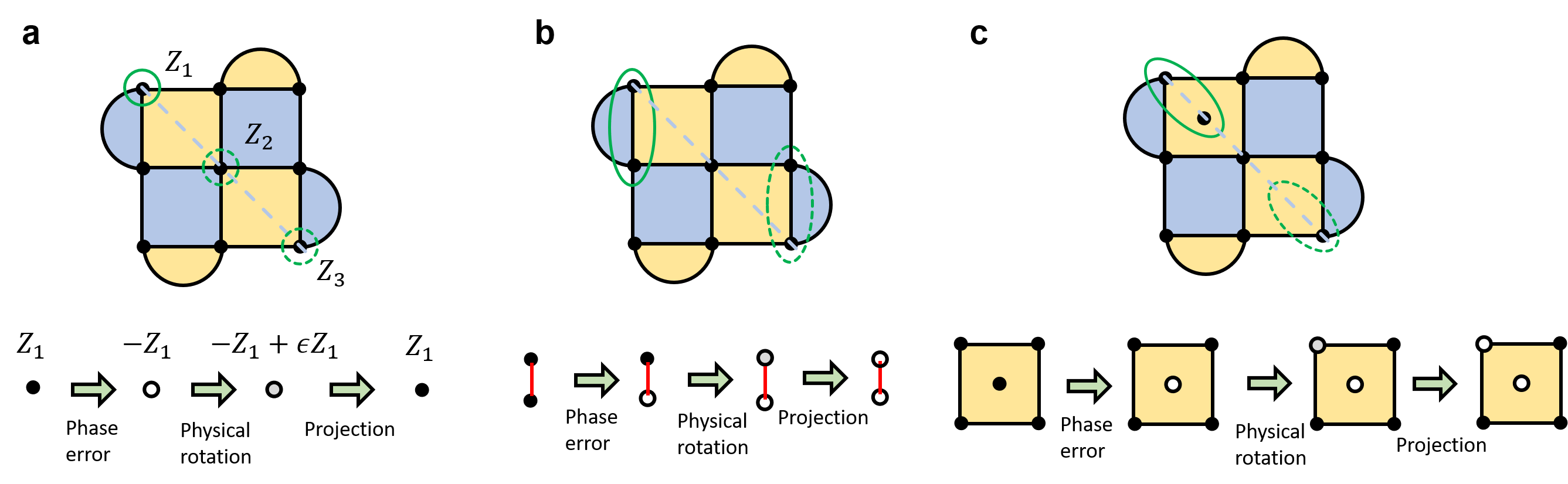}
    \caption{\textbf{Three types of errors.} \textbf{a} In a flip-projection error, a physical qubit in the logical operator (support qubit) phase-flips and is subsequently projected to a phase-flipped basis. Possible error instances are circled with green. For the surface code, there are $d$ instances. \textbf{b} A secondary-flip-projection error occurred by the combination of the phase flip of another qubit and undesirable projection of the support qubit (see the text for the detailed condition of the combination). \textbf{c} The readout error happens by the error in stabilizer measurement. The process is often modeled as the phase error (either $\hat{Z}$ or $\hat{Y}$) of the ancilla qubit. The readout errors can be easily suppressed by multiple stabilizer measurements. The qubits involved in the lowest-order readout errors are marked with green circles. There are two combinations for the surface code.}
    \label{fig:ErrType}
\end{figure*}

Our method of small-angle rotation introduces, in leading order, three types of logical errors in addition to the logical errors in the idle (Fig.~\ref{fig:ErrType}\textbf{a}). In the main text, we showed error probability with the phase-flip code example, and the logical errors from stabilizer errors are missing. In this subsection, we illustrate these errors with the surface code, including the one with the stabilizer measurement error. We emphasize that these errors are verified with the numerical simulation detailed in Appendix~\ref{app:sim}.

\begin{figure}[b]
    \centering    
    \includegraphics[width=\columnwidth]{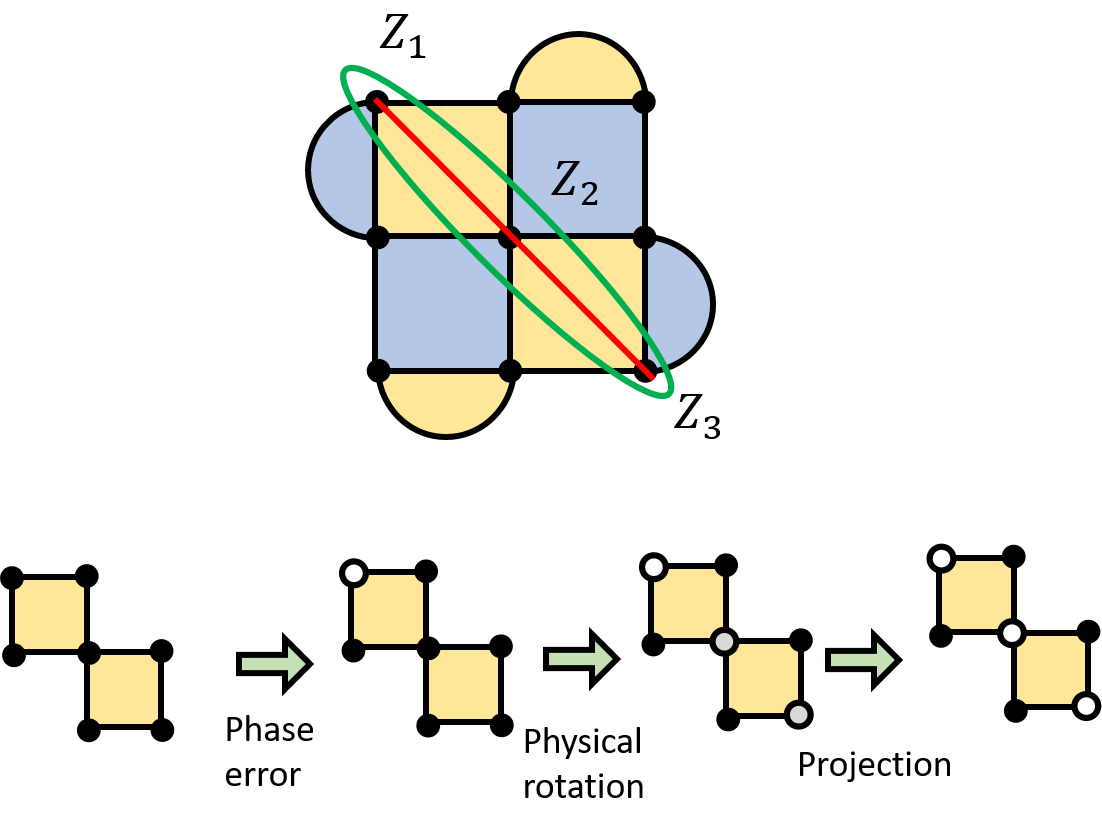}
    \caption{\textbf{String type of error.} The error happened by the combination of phase flip and undesired stabilizer projections. Though this seems a separate kind of error from Fig.~\ref{fig:ErrType}, it is a part of the flip-projection error. The string error is a small amplitude wavefunction component of flip-projection error state.}
    \label{fig:stringErr}
\end{figure}

Figure~\ref{fig:ErrType}\textbf{a} shows the flip-projection error. This error occurs when projecting one of the physical qubits on $\hat{Z}_L$ experiences a phase error, then the physical rotation is projected to the phase-flipped basis as shown diagrammatically in the lower panel. To be precise, the flip-projection error includes the wavefunction resulting from the phase-flip of one qubit and phase-flip projection of all the other qubits on the diagonal ($\epsilon^3$-component in the phase-qubit example of Eq.~\ref{eq:phaseFlipCodeError}). We showed this in Fig.~\ref{fig:stringErr}.

Figure~\ref{fig:ErrType}\textbf{b} illustrates the second type of error where one of the support qubits of $\hat{X}$ stabilizer of physically rotated qubit flipped, and that physically rotated qubit is projected to the phase-flipped basis. The resulting state does not change the sign of stabilizer measurement, and thus, is not discarded in the error-detection phase. For the surface code, there are only two cases, each with the qubits at the end of the logical string, and this is due to the topological character of the surface code. In general, the number of combinations of normalizer-supporting qubits and other qubits that are undetected by the rotation projection and the error depends on the code design. 

Figure~\ref{fig:ErrType}\textbf{c} presents the logical error by measurement errors. If the support qubit is projected to a phase-flipped basis and the stabilizer measurement is erroneous, then the logical error is undetected. This type of error can be exponentially suppressed by multiple stabilizer measurements. In our simulation, the number of stabilizer measurements is chosen to be minimal but not affect the overall error rate, to minimize the resource overhead. In our benchmark results, Pareto front points are mostly with twice of stabilizer measurements.

\section{Example: Four-qubit error detection code}\label{app:4qubit}
Four-qubit code is the smallest quantum code with $d=2$ and can detect single-qubit errors. The stabilizers of the four-qubit code is $S_z = Z_1Z_2Z_3Z_4$ and $S_x = X_1X_2X_3X_4$. It implements two logical qubits $Z_L^{(1,2)}=(Z_1Z_2,Z_2Z_3)$, $X_L^{(1,2)}=(X_2X_3,X_1X_2)$. Applying $\bigotimes_{i=1,2,3}\hat{R}_{z,i}(\theta)$ to $\ket{+_L}$ and projecting, we achieve $\theta_L = 2\sin^{-1}\left(\sin^2(\theta/2)/\sqrt{\cos^4(\theta/2)+\sin^4(\theta/2)}\right)$ rotation around $-y$ axis. Incoherent error rate for both logical qubits are $\epsilon_\text{in}^{(1,2)} = 8\frac{p_\text{in}}{3}\sin^2(\theta/2)\cos^2(\theta/2)\cos^2\theta_L$, and their errors are correlated with $\rho = \frac{1}{2}-\frac{\epsilon_\text{in}}{2(1-\epsilon_\text{in})}\approx 1/2$

For a small angle $\theta$, $\theta_L\approx \theta^2/2$ and $\epsilon_\text{in}^{(1,2)}\approx\frac{4p_\text{in}}{3}\theta_L$. One can rotate the logical qubit by $\theta_L$, $\sim1/\epsilon_\text{in}$ times with a low error. The total rotated angle is $\theta_L/\epsilon_\text{in}\approx1/p_\text{in}$, and this corresponds to $2\pi$-rotate $n$ logical qubits $1/np_\text{in}$ times. This corresponds to the order of $\sim 10$ $2\pi$-rotations for $n=100$ and $p_\text{in}=10^{-3}$, representing NISQ applications.

\section{Example: Perfect code}
The five-qubit code is the smallest QEC code with $d=3$ and is called the perfect code. The stabilizers are $S=\{XZZXI, IXZZX, XIXZZ, ZXIXZ\}$ and the logical qubit is defined to be $\{Z_L=ZZZZZ, X_L =XXXXX\}$, where we omitted subscripts for the sake of notational simplicity. Although these are conventional choices, it is beneficial for our method to express $\hat{Z}_L$ with a weight 3 Pauli product. For this purpose, we alternatively define the code with $S=\{YYIZZ, IXXXZ, YXZIX, XYZXI\}$ and $\{Z_L=ZZZII, X_L =YYXXX\}$, using the stabilizer algebra and basis change. Applying $\hat{R}_{z,L}(\theta_L) = \bigotimes_{i=1,2,3}\hat{R}_{z,i}(\theta)$ to $\ket{+_L}$ and projecting, we achieve $\theta_L = -2\sin^{-1}\left(\sin^3(\theta/2)/\sqrt{\cos^6(\theta/2)+\sin^6(\theta/2)}\right)$. The incoherent error rate for both logical qubits is $\epsilon_\text{in}= 3\frac{p_\text{in}}{3}\sin^2(\theta/2)\cos^4(\theta/2)\sin^2(\Delta\theta_L/2)$, $\Delta\theta_L = \theta_L-\theta$.

For a small angle $\theta$, $\theta_L\approx -\theta^3/4$ and $\epsilon_\text{in}\approx p_\text{in}(\theta/2)^4$. Rotating the logical qubit by $\theta_L$, $\sim1/\epsilon_\text{in}$ times. The total rotated angle is $\theta_L/\epsilon_\text{in}\approx 1/(\theta p_\text{in})$.

\section{Simulation details}\label{app:sim}
\subsection{Preliminary on subspaces of non-Clifford states}
The following three sections will provide more details on our simulation methodology for non-Clifford state preparation as well as on its complexities. Firstly, we present some basic properties of non-Clifford operators under an error detection procedure. Secondly, we introduce the extended form of stabilizer states used in our simulation algorithm. Finally, we provide the pseudocode of the simulation procedure and present results for the noisy simulation of state preparation.

We assume that the logical rotation operator $\hat{R}_{z,L}(\theta_L)$ is performed using our method in Eq.~(\ref{eq:physical_rotations}). 
We state some basic facts on stabilizer eigenspaces and their relations to our protocol. We assume, for illustration purposes, an $[[n,1,d]]$ code with logical $Z$ operator: $\hat{Z}_L = \Pi_{i=1,\dotsc, d} \hat{Z}_i$. In general, any normalizer consisting of single-qubit Paulis can follow the same analysis. Let $\ket{0_L}$ and $\ket{1_L}$ be the codewords, i.e., the simultaneous ($+1$)-eigenstates of all stabilizers $\Sigma$ of the code. The physical rotations $$U = \bigotimes_{i=1,\dotsc,d}\hat{R}_{z,i}(\theta)$$ in our method takes the quantum state out of the logical codespace, given by Eq.~(\ref{eq:physical_rotations}). We now discuss the resulting state space $\Omega$ after applying the physical rotations $U$.

\emph{Notations:} Consider a distance-$d$ surface code, we can show that there exists a set of stabilizers $S \subseteq \Sigma$ that decomposes the resulting state space into a direct sum of the $m=2^{d-1}$ two-dimensional subspaces, each associated with a unique set of eigenvalues to $S_i \in S$.  To illustrate this, we focus on the subset of stabilizers that do not commute with $U$, that is $S = \{S_i \in \Sigma: [S_i,U]\neq 0\}$, and $m=2^{|S|}$ where $|S|$ is the number of stabilizers in $S$. For example, $S$ can be the $d-1$ $X$-type stabilizers along the logical $\hat{Z}$ operator in surface code, in which case $m = 2^{d-1}$.  For some bitstring $s \in \{0,1\}^{|S|}$, let $\Omega_s$ be a space spanned by an orthonormal set of vectors $V$ such that $\ket{v} \in V$ is a $(-1)^{s_i}$-eigenstate to each stabilizer $S_i \in S$ and $(+1)$-eigenstate to all stabilizers from $\Sigma \setminus S$. For instance, if $s_i = 00\dotsb0$, then $\Omega_s$ is spanned by the codewords $\ket{0_L}, \ket{1_L}$. We now show the following direct sum decomposition $\Omega = \bigoplus_{s \in \{0,1\}^{|S|}} \Omega_s$.

For simplicity, we call the space $\Omega_s$ an $s$-eigenspace of $S$. Due to the (anti-)commutation relations of Pauli group, the dimension of $\Omega_s$ is $2^{d-|S|}$.  For the case of $|S| = d-1$, each $\Omega_s$ has dimension two. It is clear that $\Omega_s \cap \Omega_{s'} = \{\mathbf{0}\}$ for $s\neq s'$ -- assume for the sake of contradiction that there exists a nonzero vector $\mathbf{v} \in \Omega_s \cap \Omega_{s'}$, if $s$ and $s'$ differ at index $i$, then $\mathbf{v}$ must be simultaneously proportional to two orthogonal vectors, i.e., $(+1)$-eigenstate of $S_i$ and $(-1)$-eigenstate of $S_i$.

 For a bit-string $b \in \{0,1\}^d$, we define the operator 
 \begin{align}
 \hat{Z}^b = \Pi_{i: b_i=1}\hat{Z}_i    ,\label{eq:z_op}
 \end{align}
 where $\hat{Z}_i$ is the Pauli $Z$ operator on qubit $i$ along the logical $\hat{Z}_L$ operator. For example, in our surface code example, by construction, we have $\hat{I}_L = \hat{Z}^{00\dotsb 0}$ and $\hat{Z}_L = \hat{Z}^{11\dotsb 1}$. By Eq.~(\ref{eq:physical_rotations}), we have 
 \begin{align}
 U = \sum_{b \in \{0,1\}^{d}} u_b \hat{Z}^b, \label{eq:u_decomp}
 \end{align} 
 where $u_b = \cos^{d-|b|}\left(\frac{\theta}{2}\right)\left(i\sin\left(\frac{\theta}{2}\right)\right)^{|b|}$. For a codeword $\ket{x_L}$, we have the resulting state after the physical rotations as $U\ket{x_L} = \sum_b u_b \hat{Z}^b \ket{x_L}$. Denote $\bar{b}$ the bit-wise negation of the bitstring $b$ and $|b|$ the hamming weight of $b$. For a given $b$ and $|S|=d-1$, there exists a unique $s \in \{0,1\}^{|S|}$ such that $\hat{Z}^b \ket{x_L}, \hat{Z}^{\bar{b}} \ket{x_L} \in \Omega_s$. This is because $Z_L$ is a normalizer of the stabilizers $S$ (i.e., commuting with all stabilizers) and $\hat{Z}^{\bar{b}} = \hat{Z}_L\hat{Z}^{b}$, so $\hat{Z}^b \ket{x_L}, \hat{Z}^{\bar{b}} \ket{x_L}$ are simultaneous $s$-eigenstates of $S$. 
 
 Now that we have shown that the resulting state space $\Omega$ after $U$ is a direct sum of the subspaces $\Omega_s$, we can conveniently analyze the role of quantum error detection in our protocol in the next few sections. For convenience, we can write down,
 \begin{align}
    \sum_{b=0}^{2^{d}} u_b \hat{Z}^b \ket{x_L} = \sum_{b=0}^{2^{d-1}}(u_b \hat{Z}^b + u_{\bar{b}} \hat{Z}^{\bar{b}}) \ket{x_L}, \label{eq:pair_b}
\end{align}
with coefficients $u_b$, $u_{\bar{b}}$ defined in Eq.~(\ref{eq:u_decomp}).
There is a unique $s$ associated with each $b$ such that both $\hat{Z}^b \ket{x_L}, \hat{Z}^{\bar{b}} \ket{x_L} \in \Omega_s$. We also define a mapping from the bit-string $b$ to the stabilizer measurement result $s$.
$$f: \{0,1\}^d \rightarrow \{0,1\}^{|S|}$$ 
Hence, by construction, $f(b) = f(\bar{b}) = s$ if and only if $\hat{Z}^b \ket{x_L}, \hat{Z}^{\bar{b}} \ket{x_L} \in \Omega_s$.
 
\subsection{Extended form of stabilizer states}
A pure quantum state $\ket{\psi}$ is called a stabilizer state if and only if $\ket{\psi} = U_C\ket{0^n}$, for some Clifford circuit $U_C$. As shown in Eq.~(\ref{eq:u_decomp}), the non-Clifford rotation operations $\hat{R}_z(\theta)$ can be written as a linear combination of Clifford operations. In case of the magic state preparation step, we first prepare $\ket{U} = UH_L\ket{0_L}$, where $H_L$ is the logical Hadamard gate and $U$ is non-Clifford given by Eq.~(\ref{eq:physical_rotations}). By construction of the subspaces from the previous section, we can write the prepared state as a linear combination of stabilizer states. If we denote the stabilizer states $\ket{\psi_b} = \hat{Z}^b \ket{+_L}$ and $\ket{\psi_{\bar{b}}} = \hat{Z}^{\bar{b}} \ket{+_L}$, we have
\begin{align}
\ket{U} = \sum_{b=0}^{2^d} u_b \hat{Z}^b \ket{+_L} = \sum_{b=0}^{2^{d-1}}(u_b \ket{\psi_b} + u_{\bar{b}} \ket{\psi_{\bar{b}}}),
\end{align}
following Eq.~(\ref{eq:pair_b}).

In a noiseless scenario, our protocol proceeds with a round of error detection. The stabilizer measurement by $S$ yields an output string $s$. If $s \neq 00\dotsb 0$, we discard; otherwise, we obtain a ``$00\dotsb 0$''-eigenstate of $S$. Therefore, after post-selection, the state is projected to (up to normalization),
\begin{align}\ket{U} \xrightarrow{QED} u_b \ket{\psi_b} + u_{\bar{b}} \ket{\psi_{\bar{b}}}, \label{eq:subspace}
\end{align}
where $b,\bar{b}$ satisfy $f(b)=f(\bar{b}) = 00\dotsb 0$.

If, however, some error $E$ (sampled from depolarizing channel noise) occurs during the physical rotations $U$, our protocol could yield a different outcome. Some errors, such as bit flip errors on data qubits, will be detected as usual, while others, such as phase errors on any data qubits and bit errors on ancilla qubits, can pass through error detection and manifest in the resulting state. Fortunately, we can model their effects on the resulting state by an unknown shift in the stabilizer measurement mapping. Specifically, we denote the string $e \in \{0,1\}^d$ induced by the error $E$. Then the noisy state $\ket{U_{E}}$ will be projected to
\begin{align}\ket{U_{E}} = E\ket{U} \xrightarrow{QED} u_{b'} \ket{\psi_{b'}} + u_{\bar{b'}} \ket{\psi_{\bar{b'}}}, \label{eq:noisy_subspace}
\end{align}
where $b',\bar{b'}$ satisfy $f(b'\oplus e)=f(\bar{b'}\oplus e) = 00\dotsb 0$. Here, $\oplus$ is the bit-wise xor operation.

As such, the resulting quantum state of QED (i.e., the ``$00\dotsb 0$''-eigenstate of $S$) in the presence of noise can be written as a linear combination of two stabilizer states $\ket{\psi_{b'}}, \ket{\psi_{\bar{b'}}}$. We can uniquely determine the state and its error detection outcome with a tuple, $(b', e)$, representing the projected rotation operator $\hat{Z}^{b'}$ and the error string $e$, respectively. The resulting logical rotation as a function of $b'$ is given by:
\begin{align}
    \varphi_L(b') = 2\sin^{-1}\left(\frac{u_{\bar{b'}}/i}{\sqrt{u_{b'}^2 + u_{\bar{b'}}^2}}\right) \label{eq:generic_logical}
\end{align}
where $u_{b'} = \cos^{d-|b'|}\left(\frac{\theta}{2}\right)\left(i\sin\left(\frac{\theta}{2}\right)\right)^{|b'|}$ and $\theta$ is the physical rotation angle. As such, it gives us a means to efficiently simulate the noisy non-Clifford state preparation process using Monte-Carlo method. We describe the simulation algorithm in the next section.

\subsection{Simulation of noisy state preparation}

We now describe an extension to the stabilizer-based simulation to calculate the output states from our state preparation procedure (Algorithm~\ref{algo:sim}) and estimate the logical fidelity of the output state (Algorithm~\ref{algo:fidelity}).

\begin{algorithm}[H]
  \caption{Extended Stabilizer Simulation} \label{algo:sim}
   \begin{algorithmic}[1]
   \State $p_{\text{in}} \gets \text{parameter for depolarizing channel}$
   \State $r \gets \text{error detection cycles}$
   \Procedure{NoisyStatePreparation}{$\theta_L, \theta$} 
   \State \algorithmiccomment{To prepare  $\ket{M_{\theta_L}}$ in Eq.~(\ref{eq:magic_state}) using physical rotations $\hat{R}_z(\theta)$}
   \State $\alpha_I \gets \cos(\theta/2)$, $\alpha_Z \gets \sin(\theta/2)$
   \State $d \gets \text{code distance}$
   \State Initialize stabilizer state $\ket{+_L}$ 
   \State Sample $b \in \{0,1\}^d$ where $b_i\sim Bernoulli(\alpha_Z^2)$
   \For{$k \gets 1 \dotsc r$}
       \State Sample errors $E$ from $\mathcal{E}_{p_{\text{in}}}$ using Eq.~(\ref{eq:depolar})
       \State $\ket{\psi} \gets E\hat{Z}^b\ket{+_L}$ using Eq.~(\ref{eq:z_op})
       \If{$\ket{\psi} \notin \Omega_s$ where $s=00\dotsb 0$}
           \State \Return (Failure, $b$) \algorithmiccomment{Discard if error detected}
       \EndIf
   \EndFor
   \State \Return (Success, $b$) 
   \EndProcedure
   \end{algorithmic}
\end{algorithm}
\;
\begin{algorithm}[H]
  \caption{Logical Fidelity Estimation}\label{algo:fidelity}
   \begin{algorithmic}[1]
   \State $N \gets \text{number of samples}$
   \Procedure{FidelityEstimate}{$\theta_L, \theta$} 
   \State $\alpha_I \gets \cos(\theta/2)$
   \State $\alpha_Z \gets \sin(\theta/2)$
   \State $d \gets \text{code distance}$
   \State $\mathcal{E} \gets \O$
   \For{$k \gets 1 \dotsc N$}
        \State $(e, b) \gets$ NoisyStatePreparation$(\theta_L, \theta)$
        \If{$e$ is Success}
            \State $m \gets \min(|b|, d-|b|)$
            \State $\varphi_L \gets \varphi_L(m)$ given by Eq.~(\ref{eq:generic_logical})
            \State $\mathcal{E} \gets \mathcal{E} \cup \left\{(\sin((\theta_L-\varphi_L) / 2))^2\right\}$
        \EndIf
   \EndFor
   \State Compute $\mathbb{E}[\mathcal{E}]$
   \EndProcedure
   \end{algorithmic}
\end{algorithm}

We assume a depolarizing channel, 
\begin{align}
\mathcal{E}_{p_{\text{in}}}(\rho) = (1-p_{\text{in}})\rho + (p_{\text{in}}/3)(X\rho X + Y\rho Y + Z\rho Z) \label{eq:depolar}    
\end{align}
with input error probability parameter $p_{\text{in}}$. The error channel is applied to all physical qubits (both data and ancilla qubits) during each error detection cycle. In our experiments from Figure~\ref{fig:SimResult}, $p_{\text{in}}$ takes value $10^{-3}$, representing realistic near-term noisy devices.

The main reason that we can use a stabilizer-based simulator for a non-Clifford state preparation circuit is that our circuit contains only one non-Clifford gate, namely a logical $R_z$ rotation, which can be decomposed into a linear combination of $\{I, Z\}^{\otimes d}$ as shown in Eq.~(\ref{eq:physical_rotations}). While the number of terms in the liner combination can grow to $2^d$, we only need to keep track of two of them. Specifically, the stabilizers in an error correction cycle will project to the subspace of only two operators, i.e., $\hat{Z}^b$ and $\hat{Z}^{\bar{b}}$ satisfying $f(b) = f(\bar{b}) = 0$ as defined in Eq.~(\ref{eq:subspace}). Therefore, we can simulate the projected noisy state (in Eq.~\ref{eq:noisy_subspace}) by sampling the $n$-bit string $b$ and sampling errors $E$ from $\mathcal{E}_{p_{\text{in}}}$. The procedure for simulating noisy non-Clifford state preparation is described in Algorithm~\ref{algo:sim}. The success rate of the resulting state can be estimated with Monte-Carlo simulations. Line 12 of Algorithm~\ref{algo:sim} performs one round of error detection. 
The logical fidelity of the resulting state can be estimated using Algorithm~\ref{algo:fidelity}. The resulting logical rotation angle is given by Eq.~(\ref{eq:generic_logical}). 

\section{Clifford Cost of Angle Synthesis}\label{app:Clifford}

In the calculation of SR and COH method in Fig.~\ref{fig:resource}, we only considered the spacetime cost of magic state ancillae fed into the circuit. To implement a logical rotation gate, $R(\theta_L)$, with these ancillae, one needs many Clifford gates. For example, we calculate the physical Clifford gate costs for implementing logical $H$, $S$ and $T$ gates. We define the Clifford cost of a logical gate in the following context,
\begin{align}
    \text{Gate cost} &= \text{  Ancilla cost  } + \text{Clifford cost}. \\
    & \text{  (pre-distillation)  } \nonumber
\end{align}

Such gate cost is highly dependent on the substrate error correction code and the choice of physical implementations of the logical gates. In this Appendix, we give a detailed analysis of the costs of Clifford operations in Selinger-Ross' and Campbell-O'Gorman-Howard's methods.

\subsection{The costs of H, S, and T gates}

Implementing $S$ and $T$ gates on a logical qubit of surface code requires an ancilla state, whereas the $H$ gate can be transversally implemented with a spacetime cost of $d^3$. We assume that the ancilla for $S$ has zero cost because the ancilla state for the $S$ gate, $\ket{Y}$, can be reused. We use Gidney's ``slightly smaller $S$ gate'' with a Clifford depth of 3~\cite{gidney2017slightly}. The ancilla state should be online during the entire circuit, and the cost of an $S$ gate is $6d^3$.

On the other hand, $T$ teleportation requires a CNOT gate and an $S$ correction depending on the measurement of the ancilla. Since the $S$ fix-up only applies with a probability of 1/2, we set the Clifford cost of $T$ as $2d^3 + 0.5\times6d^3 = 5d^3$. Note that we ignored the cost of logical measurement, which takes one cycle, compared to the other Clifford operations that take $d$ cycles (measuring all physical qubits of a surface-code logical qubit $d$-redundantly measures the logical Pauli's).

\begin{table}[t!]
    \centering
    \begin{tabular}{||c|c|c|c||}
    \hline
    \hline
         Logical Gate & $H$ & $S$ & $T$  \\\hline
         Clifford Cost & $d^3$ & $6d^3$ & $5d^3$ \\
    \hline
    \hline
    \end{tabular}
    \caption{Clifford costs of single qubit gates}
    \label{tab:singleQubitCliffordCost}
\end{table}

\begin{table}[t!]
    \centering
    \begin{tabular}{||c|c|c|c|c||}
    \hline
    \hline
    Logical Angle & $H$ Count & $S$ Count & $T$ Count & Clifford Cost \\
    \hline
         $\theta_L = 2\pi/2^4$ & 15 & 2 & 14 & $97d^3$ \\
         $\theta_L = 2\pi/2^7$ & 23 & 2 & 22 & $145d^3$ \\
         $\theta_L = 2\pi/2^{10}$ & 30 & 1 & 30 & $186d^3$ \\
    \hline
    \hline
    \end{tabular}
    \caption{Gate counts and Clifford costs of SR implementation}
    \label{tab:SRClifford}
\end{table}

\begin{table}[t!]
    \centering
    \begin{tabular}{||c|c||}
    \hline
    \hline
         Pivotal rotation success & $165d^3$ \\
         Pivotal rotation fail & $181d^3$ \\
         State prep. (injection) & $6d^3$\\
         Reuse 2$\theta$ & $-1d^3$ \\
         Raw $T$ injection & $8d^3$ \\
         \hline
         Average cost (w/o $2\theta$ reuse) & $187d^3$ \\
    \hline
    \hline
    \end{tabular}
    \caption{Clifford costs of COH implementation}
    \label{tab:COHClifford}
\end{table}

Table~\ref{tab:singleQubitCliffordCost} summarizes the results of the Clifford costs in $H$, $S$, and $T$ gate applications.

\subsection{Selinger-Ross method}

We obtained the sequence of decomposed rotation gates using the Gridsynth~\cite{selinger2013gridsynth}. We reduced the Gridsynth results with $ST = TS =T^\dagger$. The numbers of $H$, $S$, and $T$ as well as the total Clifford cost in the resulting sequence are shown in Table~\ref{tab:SRClifford}. The Clifford cost of a logical rotation increases for higher-order Clifford gates. 

\subsection{Campbell-O'Gorman-Howard method}

The Campbell-O'Gorman-Howard (COH) method has two different implementations. One uses $8$ $T$ gates, as described in \cite{campbell2016efficient}. The other implementation utilizes a $CCZ$ gate, which can also be synthesized with $8$ $T$ gates~\cite{campbell2018magic}. The first version of the method was developed as an optimized version of Duclos-Cianci and Poulin's method~\cite{duclos2015reducing}, which is based on Meier, Eastin, and Knill's distillation technique~\cite{meier2012magic}. The second version of the method performs parity checking with $CCZ$, and it can be made even more efficient by combining multiple rotations with synthilation~\cite{campbell2017unifying}. We consider the second version of the method because it has a lower Clifford cost implementation, but we do not consider synthilated higher order parity check because teleportation of pivotal rotations only probabilistically succeeds (see below).

COH parity checker consumes one $2\theta$ rotation, called pivotal rotation, and a $CCZ$ gate to purify two ancillae corresponding to the $\theta$ rotation ($R_z(\theta)\ket{+}$). The pivotal rotation gate cannot be applied directly and needs to be teleported. The teleportation only succeeds with a probability of 1/2, and with the other probability of 1/2, $R_z(-2\theta)$ is teleported. To correct this, we can try again probabilistically teleporting $R_z(4\theta)$ with the $R_z(4\theta)\ket{+}$ ancilla. Successive corrections of a pivotal rotation increase logical errors, and the ancillae for correction are similarly as expensive as the initial ancillary rotation for the pivotal rotation (relative costs of each ancilla depend on the implementation, but this is confirmed from the simulation result as shown in Fig.~\ref{fig:resource}). Thus, instead, we assume aborting the process when the teleportation of pivotal rotation fails and repeating from the start. Table~\ref{tab:COHClifford} shows the Clifford cost of COH circuit in case of success and failure of pivotal rotation.

The circuit overhead above did not include the state preparation and injection costs. Total $6d^3$ state preparation and $8d^3$ of $T$ injection cost are required. When pre-distilled $2\theta$ rotation ancilla or $T$-magic state is used, $d^3$ per state should be subtracted as shown in Table~\ref{tab:COHClifford}.

\end{document}